\documentclass{appolb}
\usepackage{epsfig}
\usepackage[latin2]{inputenc}
\newcommand{\be}{\begin{equation}}
\newcommand{\ee}{\end{equation}}

\begin{document}

\title{ Correlations between the most developed (G7) countries. A moving average
window size optimisation. \thanks{Presented at First Polish Symposium on Socio and
Econophysics, Warsaw 2004} }
\author{Janusz Miskiewicz
\address{Institute of Theoretical Physics, Wroclaw University, pl. M. Borna 9, \\
50 -204 Wroclaw, Poland } \and
Marcel Ausloos
\address{ SUPRATECS, B5, University of Li$\grave e$ge, B-4000 Li$\grave e$ge,
Euroland } }
\maketitle
\begin{abstract}
Different distance matrices are defined and applied to look for correlations between the gross domestic product of G7 countries. Results are illustrated through displays obtained from various graph methods. Significant similarities between results are obtained. A procedure for choosing the best distance function is proposed taking into account the size of the window in which correlation are averaged.
\end{abstract}

\section{Introduction}

Various conclusions on correlations  depend on the window size in which the averaging technique is performed, e.g. one can obtain correlation lengths, Hurst exponents, detrended fluctuation analysis exponents, \cite{hurst,hursta,window,dfa} etc. There are two main competitive factors: requirements on statistical precision and information loss. If the time window size is large the "quality" of calculated statistical parameters is high (due to the central limit theorem in probability theory \cite{probab}) while the information is lost and the results are less sensitive to local features if the window is small. In the case of economy and financial time series it is known that the time series are nonstationary, so not only the value of considered parameters evolves in time but also their stochastic properties. Therefore the problem of the time window size is one of the very important factors in such analyses. 

There is also another factor for optimising the choice of the time window size: numerical stability requirements. In the case of any numerical calculation  every step of the procedure introduces an error. Numerical errors accumulate and in some cases (especially in nonlinear analysis) can quickly  influence the results. Therefore this factor should be taken under consideration when deciding upon the size of the time window. Therefore the procedure of adjusting the size of the time window is an optimisation problem with  two opposite competing factors. In order to find correlations  the time window must be moved along the signal, in so  doing the subsequent analysis is a ''moving average'' size window optimisation.

The analysis of the time window optimisation is herebelow done on the basis of  correlation analyses between G7 countries (France, USA, United Kingdom, Germany\footnote{Germany is considered as a one country. To have a record before consolidation the data are constructed as a sum of GDP of both German countries.}, Japan, Italy, Canada). The macroeconomy situation is described by theirs Gross Domestic Product (GDP), since in most countries GDP is considered as
an official parameter of the economic situation. GDP is usually defined as a sum of all final goods and services produced in the country, i.e. equal to the total consumer, investment and government spending, plus the value of exports, minus the value of imports\footnote{http://www.investorwords.com/2153/GDP.html}. Additionally in order to define a reference country an artificial "All" country is constructed. GDP of "All" country is defined as a sum of GDP of all 7 countries. So the GDP
increment of "All" can be considered as an averaged level of development.  

The GDP values for each of these countries are first normalised to their 1990 value given in US dollars as published by the Groningen Growth and Development Centre on their web page\footnote{http:$ \slash \slash$www.ggdc.net$ \slash
$index-dseries.html\#top}. The data cover the period between 1950 and 2003, i.e. 54 points for each country.  

\section{Distance definition}
\label{definition}

The equal time $t$ correlation function between $A$ and $B$ is defined as 

\begin{equation}
\label{korel}
\begin{array}{ll}
corr_{(t,T)} (A,B) = &
\frac{<A B>_{(t,T)} - <A>_{(t,T)} <B>_{(t,T)} }{\sqrt{(<A2>_{(t,T)} -
<A>^2_{(t,T)}) (<B2>_{(t,T)} - <B>^2_{(t,T)}})} . \end{array}
\end{equation}
The brackets $ < \ldots > $ denote a mean value over the time window $ T $ at time $ t $.  In the following, $A$ and $B$ will be the GDP yearly increments of a given country, i.e.   
\be
\label{increm}
\Delta GDP(t) = \frac{GDP(t) - GDP(t-1)}{GDP(t-1)}. 
\ee

Since the time series consists of a discrete set of numbers and the time evolution of which is thought to be stochastic the following metrics are used and the results compared. 
\begin{enumerate}
\item The statistical distance  $d(A,B)_{(t,T)} $ is 
\begin{equation}
\label{stat}
d(A,B)_{(t,T)} = \sqrt{\frac{1}{2}(1- corr_{(t,T)} (A,B))}, 
\end{equation}
where $ t $ and $ T $ are the final point and the size of the time window over which an average is taken respectively. 
\item The discrete distance  $d_{L_q} (A,B)$ is defined as the sum of absolute values between time series, i.e.
\begin{equation}
\label{discrete}
d_{L_q} (A,B) = ( \sum_{i=1}^n |a_i - b_i |^q)^{\frac{1}{q}} , 
\end{equation}
where $ A,B $ are time series: $ A=(a_1,a_2,\ldots , a_n) $, $ B=(b_1,b_2, \ldots , b_n) $. 
\item The distribution distance $d_{{\cal L}_q} (A,B)$ the distance defined between distribution functions. As a initial step a distribution function should be chosen on the basis of statistical tests, then the considered distribution functions have to be fitted (or appropriate parameters calculated). Since the statistical parameters describing GDP increments are very close to the normal distribution \cite{bidirectional} it is hereby assumed that the GDP increments are truly described by the normal distribution. The distance is taken as the metrics of ${\cal L}_q$ in Hilbert space \cite{analiza}, i.e.
\begin{equation}
\label{distr_dist}
d_{{\cal L}_q} (A,B) = [\int_{-\infty}^{+\infty} |p_A (r) - p_B(r )|^q dr]^{\frac{1}{q}} , 
\end{equation}
where $p_A(r)$ and $p_B(r)$ are the appropriate distribution function fitted to the data. 
\end{enumerate}

For the sake of result clarity the analysis is restricted to the case of $q=1$. The properties of distances measures with $q>1$ will be discussed elsewhere.

There are different advantages and inconveniences to those distance functions. Eq.(\ref{stat}), a statistical distance, is specially sensitive to observing linear correlations. The discrete Hilbert space $ L_q $ distance, Eq.(\ref{discrete}), can be applied to any data and does not require any special properties of the data, thereby seeming to be very useful for comparing various sets of data. The distribution distance, Eq.(\ref{distr_dist}), is the most sophisticated one since it requires a knowledge of the data distribution function, but it allows to compare the statistical properties of the data. The main disadvantage of this method is its sensitivity to the size of the data set, since it compares the (assumed) distribution functions. 

\section{Network definition}

In order to obtain the information about the correlations between countries a graph analysis of distance matrices is performed. The described below graphs (LMST, BMLP and UMLP) are built as a function of time and for moving time windows of various sizes (from 5 years (yrs) up to 52 yrs). The size of the time window is constant during the displacement. The mean distance between countries is calculated and averaged over number of generated graphs. (The number of generated graphs is equal to the difference of increments data points (here 53) and the size of the time window.) Finally statistical properties (mean value, standard deviation, skewness and kurtosis) of the twice averaged distance between countries are calculated and discussed. Within the paper the mean value of the distance between countries is understood as a mean value averaged over the number of links on a graph and the number of calculated graphs. 

The graph analysis of the distance matrices are based on three types of graph structures:
\begin{description}
\item[LMST] The Locally Minimal Spanning Tree is the version of a Minimal Spanning Tree (MST) under the assumption that the root of MST is the pair of closest neighbours.
\end{description}
For the sake of simplicity  the minimal length path algorithm (MLP) \cite{bidirectional} is used. It is a 1-D modification of the MST algorithm. This algorithm emphasises the strongest correlation between entities with the constraint that the item is attached only once to the network. This results in a lack of loops on the ''tree''. Two different graphs: the unidirectionally  growing and the bidirectionally growing minimal length paths (UMLP and BMLP respectively) are constructed. The UMLP and BMLP algorithms are defined as follows:
\begin{description}
\item[UMLP] The algorithm begins with choosing an initial point of the chain. Here the initial point is the "All" country. Next the shortest connection between the initial point and the other possible neighbours (in terms of the distance definition - Eq.(\ref{discrete}), (\ref{stat}) or (\ref{distr_dist}) is looked for. The closest possible one is selected and the country attached to the initial point. One searches next for the entity closest to the previously attached one, and repeats the process. 
\item[BMLP] The algorithm begins with searching for the pair of countries which has the shortest distance between them. Then these countries become the root of a chain. In the next step the closet country for both ends of the chain is searched. Being selected it is attached to the appropriate end. Next a search is made for the closest neighbour of the new ends of the chain. Being selected, the entity is attached, a.s.o.  
\end{description}

\section{Distance and network analysis}

\begin{figure}
	\includegraphics[angle=-90,scale=0.25]{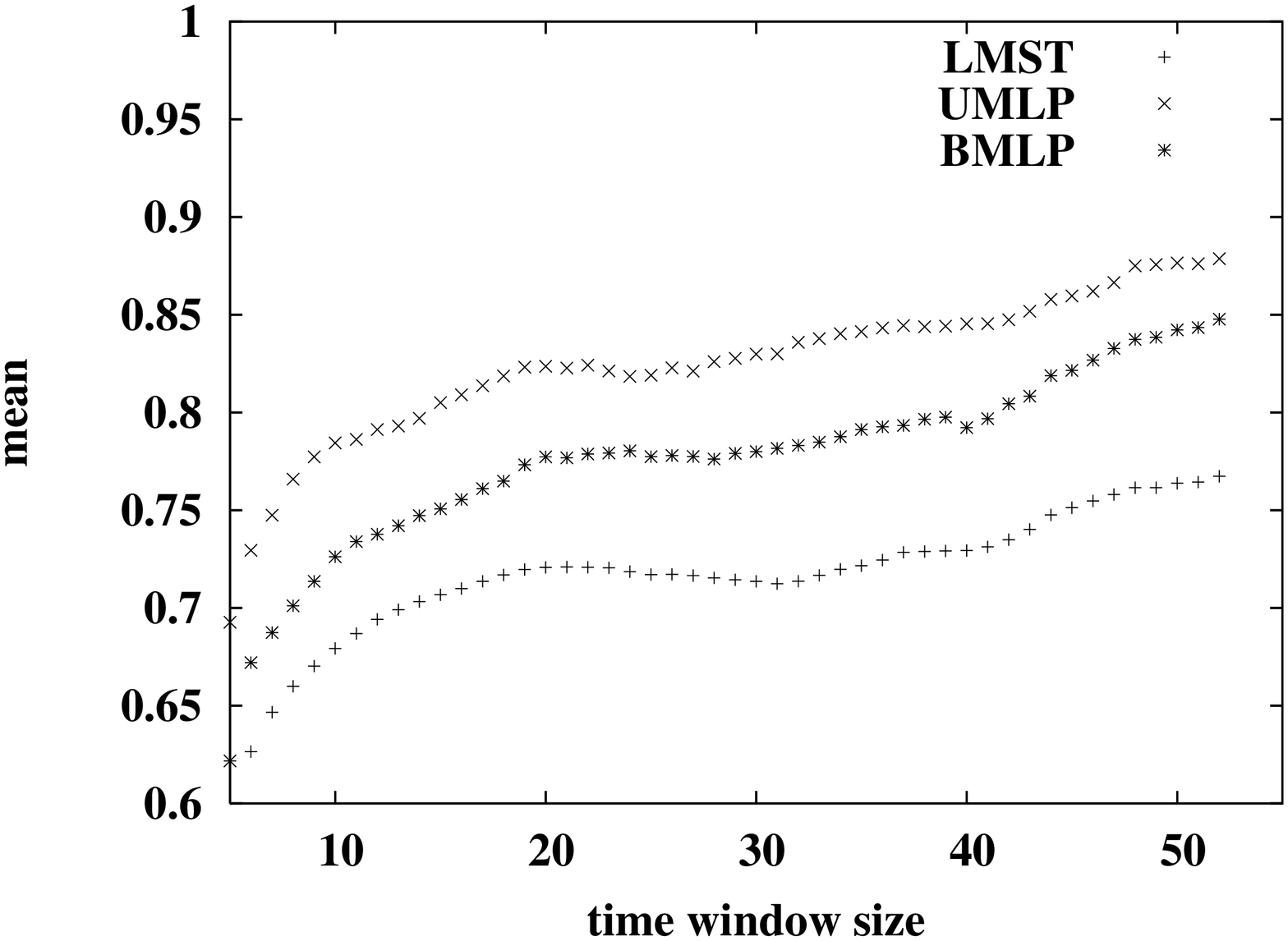}
	\includegraphics[angle=-90,scale=0.25]{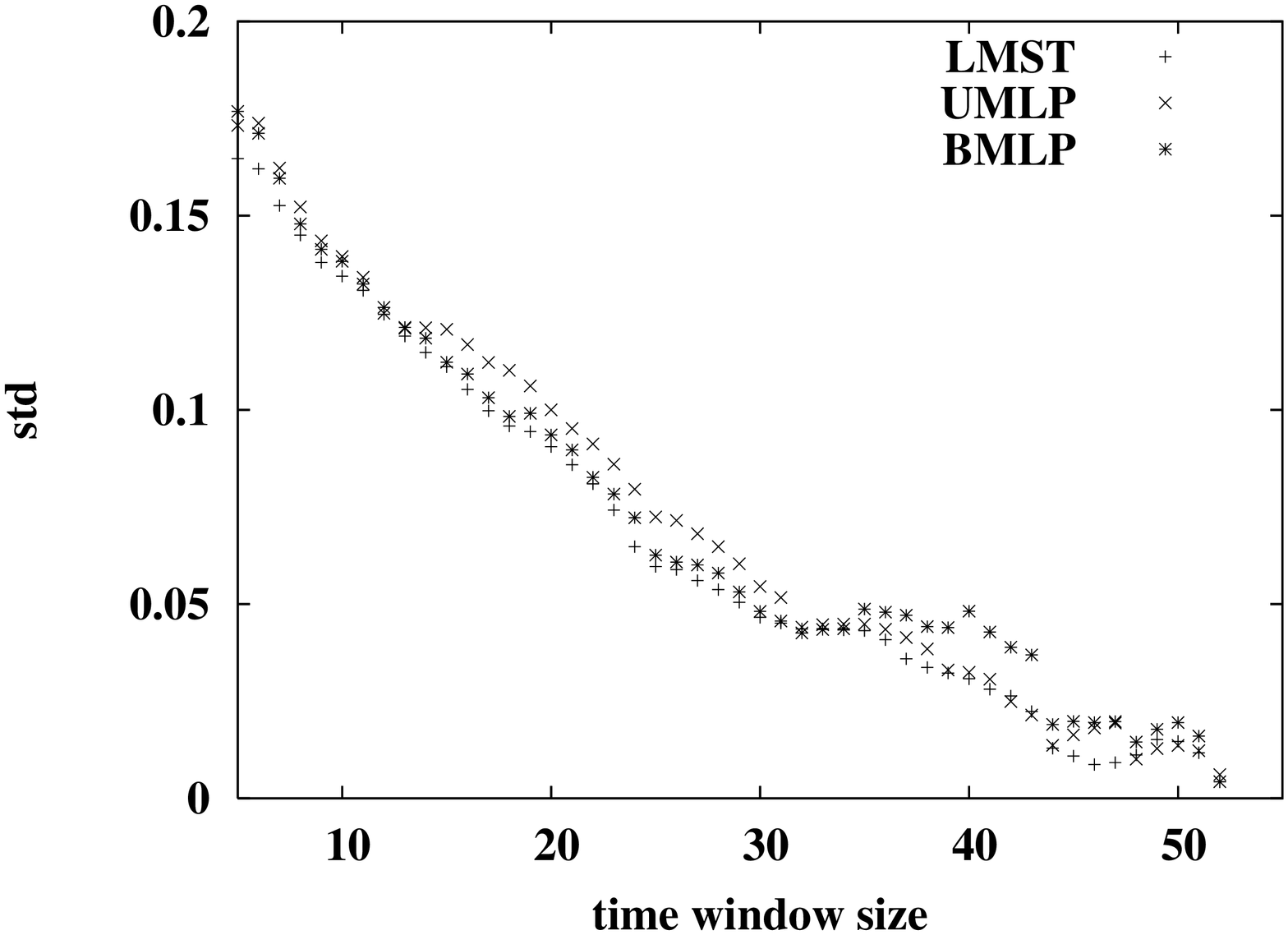} 
	\includegraphics[angle=-90,scale=0.25]{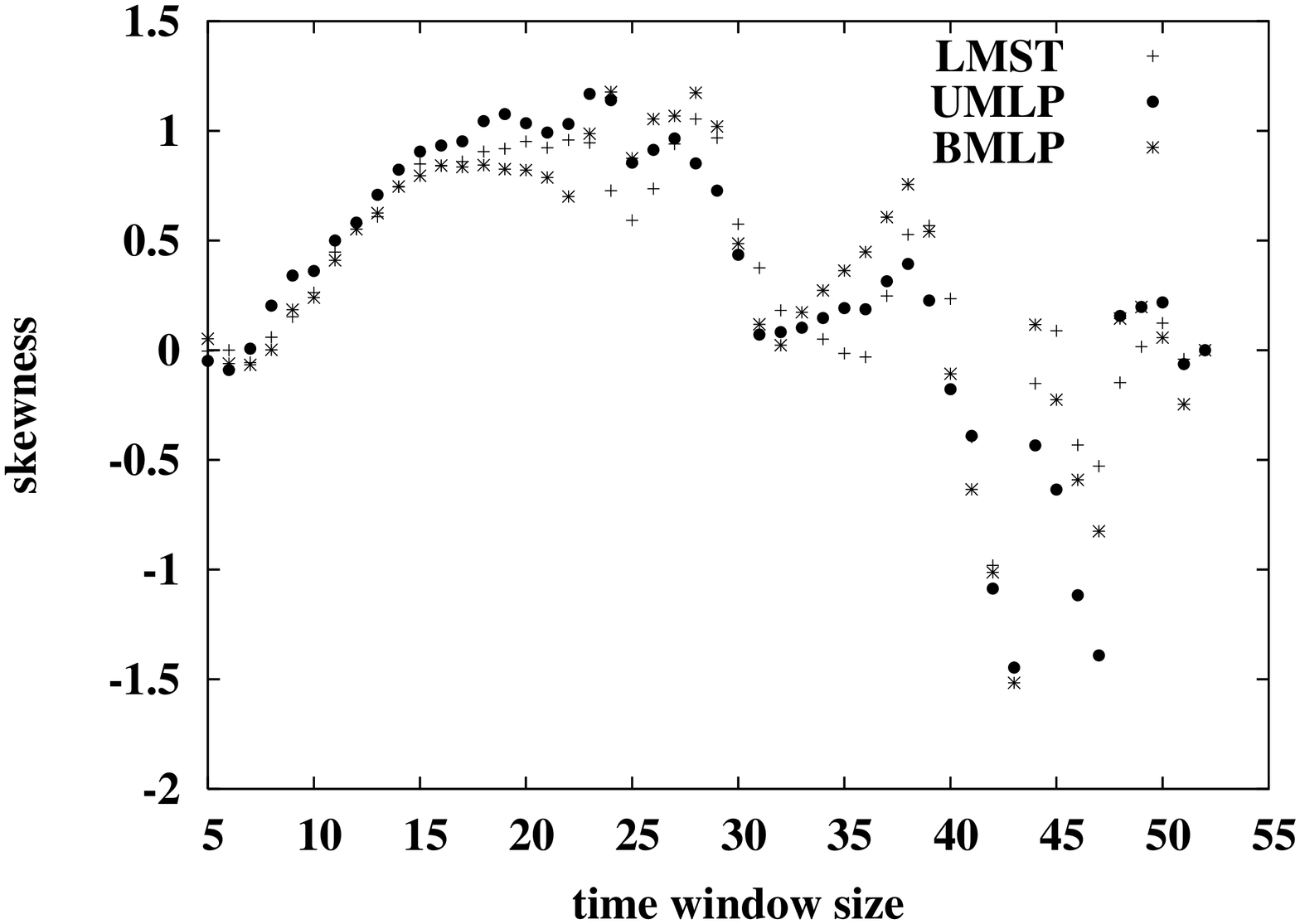}
	\includegraphics[angle=-90,scale=0.25]{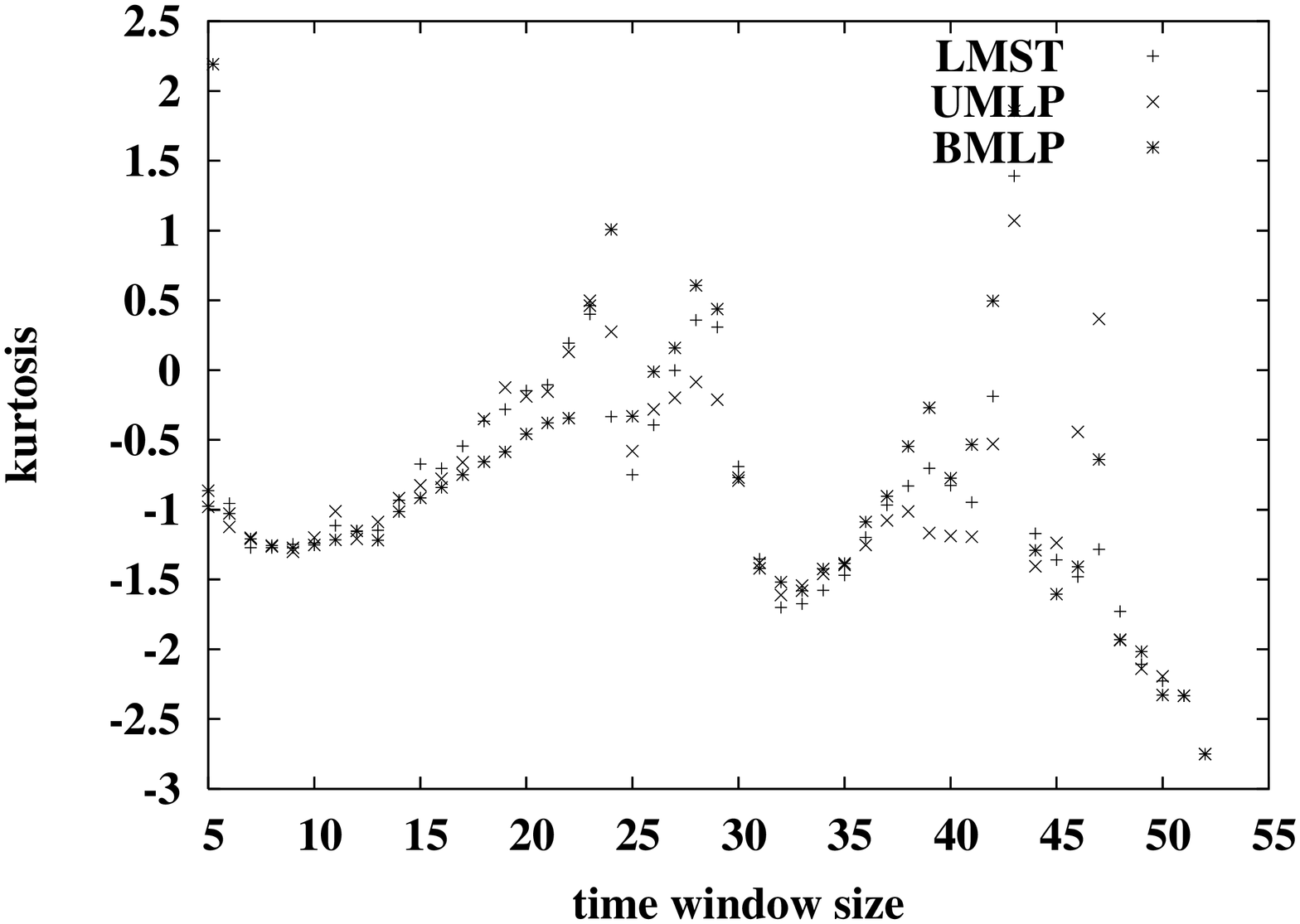}
	\caption{Statistical analysis of the graph properties obtained by application
of the statistical distance. The plots present mean value, standard deviation,
skewness and kurtosis of the distance between countries in the case of LMST, UMLP
and BMLP averaged over all G7 countries and the considered time interval as a
function of the time window size.} 	\label{fig:m}
\end{figure}

\begin{figure}
	\includegraphics[angle=-90,scale=0.25]{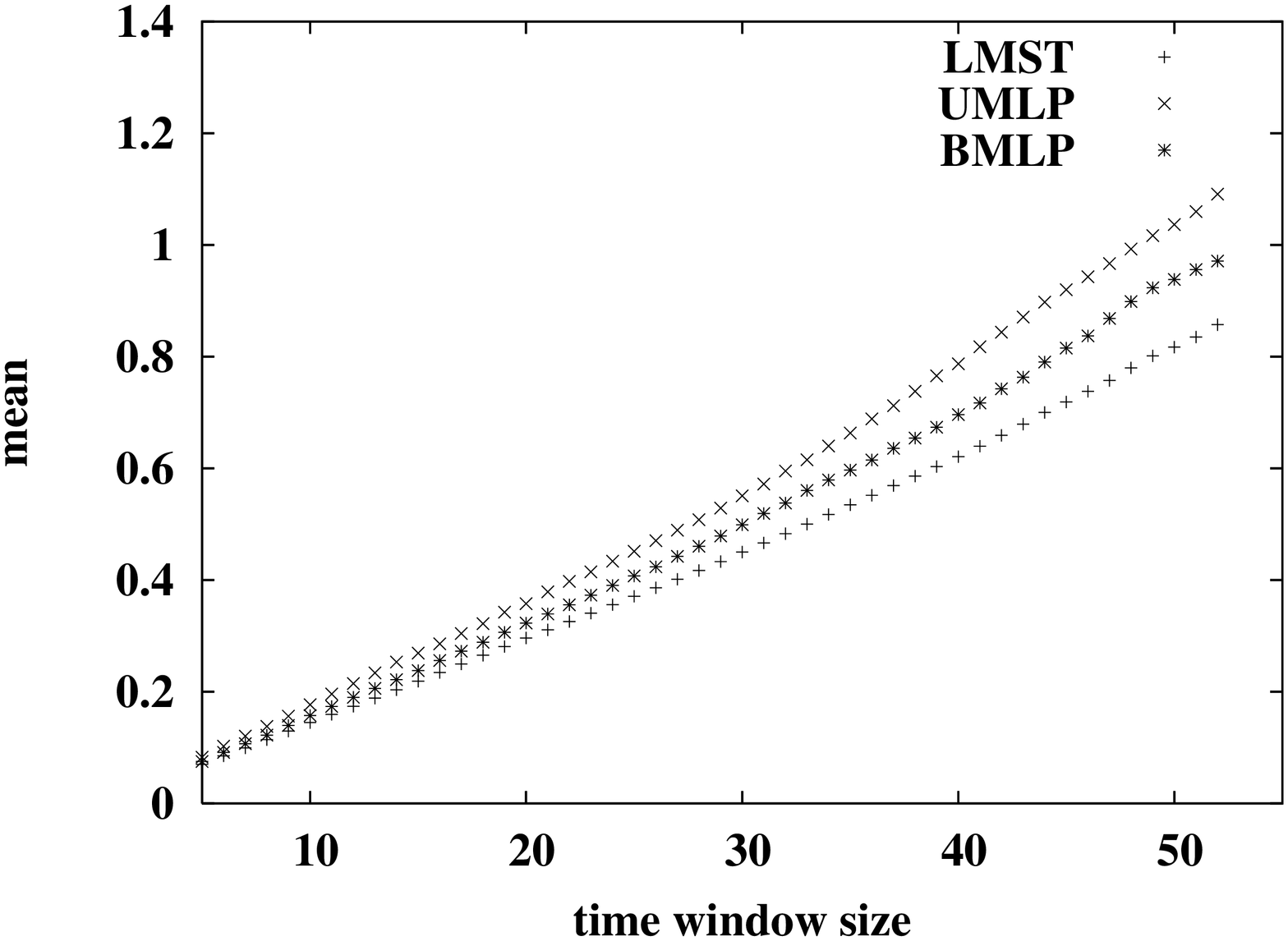}
	\includegraphics[angle=-90,scale=0.25]{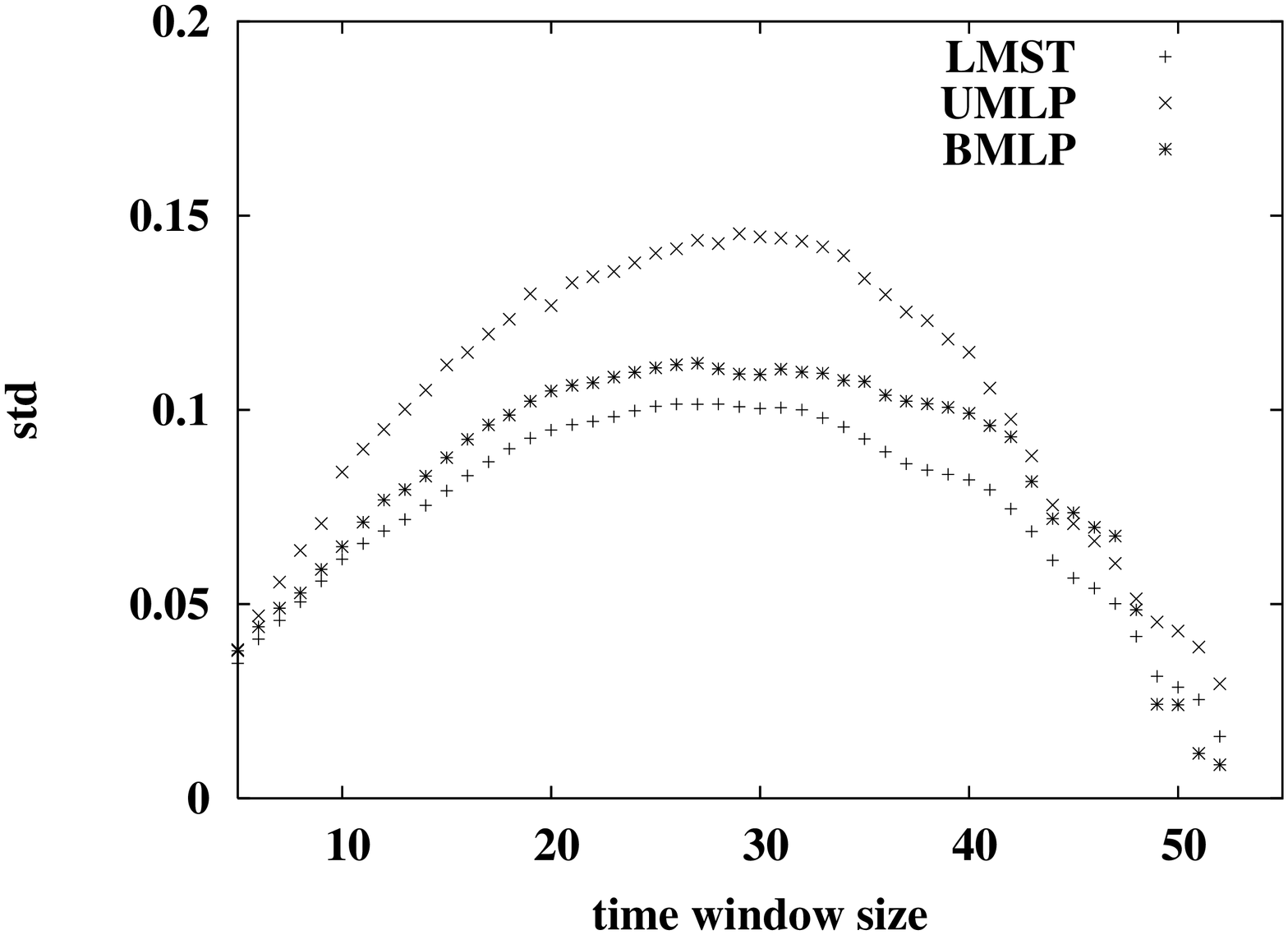} 
	\includegraphics[angle=-90,scale=0.25]{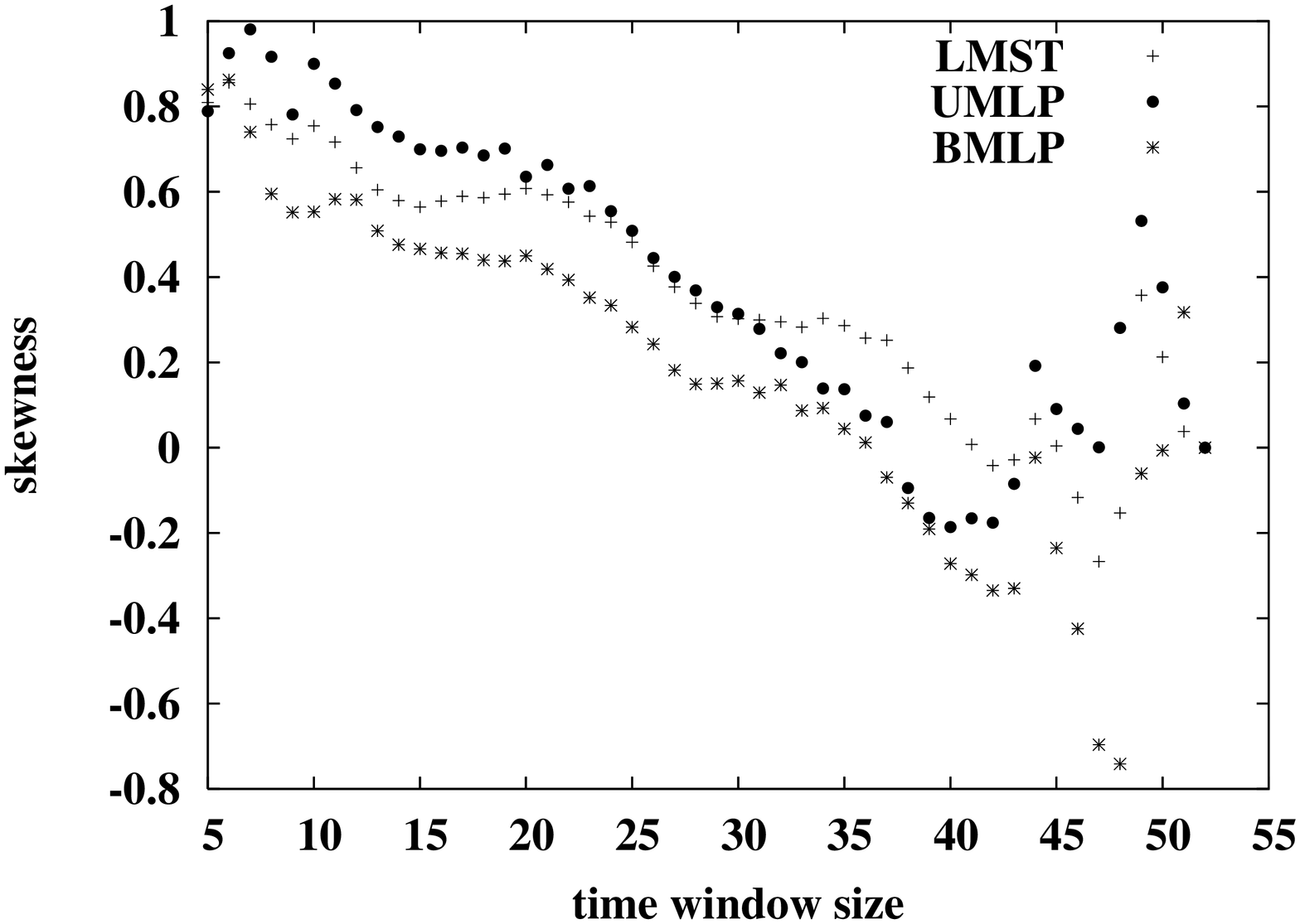}
	\includegraphics[angle=-90,scale=0.25]{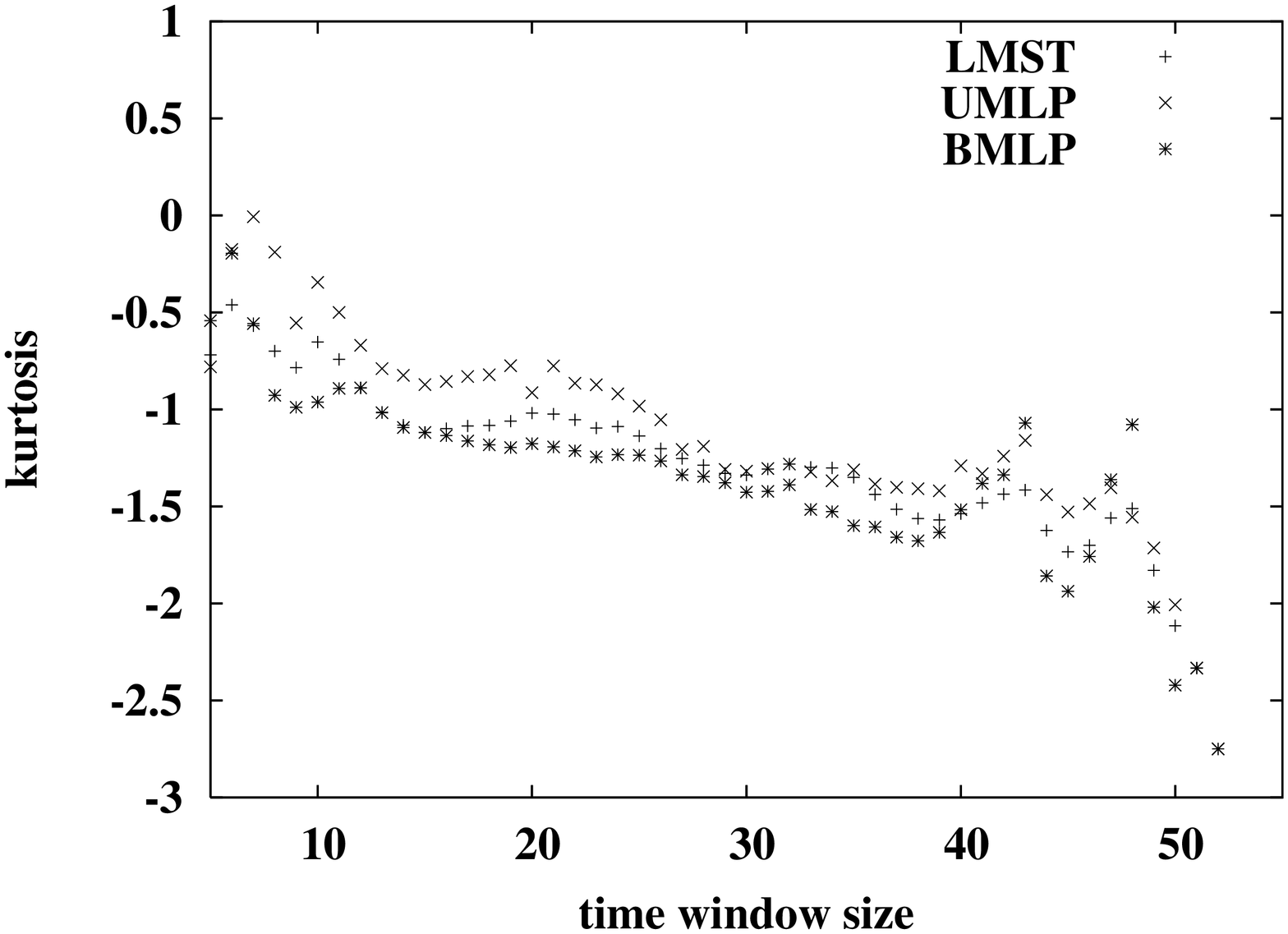}
	\caption{Statistical analysis of the graph properties obtained by application
of the $L_1$ distance. The plots present mean value, standard deviation, skewness
and kurtosis of the distance between countries in the case of LMST, UMLP and BMLP
averaged over all G7 countries and the considered time interval as a function of
the time window size.} 	\label{fig:l1}
\end{figure}

The results are presented for every distance measure, i.e. application of statistical distance in Fig.\ref{fig:m}, discrete distance in Fig.\ref{fig:l1} and finally the distribution distance in Fig.\ref{fig:m}. 

In the case of the statistical distance (Eq.(\ref{stat})) and the discrete metrics (Eq.(\ref{discrete})) (Fig.\ref{fig:m} and Fig.\ref{fig:l1} respectively) the mean distances (understood as it is defined in the section \ref{definition}) between considered countries increase with the time window size.

% Mantegna

The results obtained by application of the statistical distance, Eq.(\ref{stat}) is are presented in the Fig.\ref{fig:m}.
The averaged distances between countries are very similar in all considered graph methods and almost parallel to each other. In the case of other basic statistical properties the similarities are even stronger. For other considered statistical parameters i.e. standard deviation, skewness and kurtosis plots are almost identical. The standard deviation is decreasing (except for a few points). This suggests that linear correlations between countries are better seen for longer window size and the co-operation between those countries has a stable form best seen in the long time scale. Of course there are problems with the size of considered data. While increasing the time window size the amount of data is decreasing which results in significant changes in skewness and kurtosis value for time window size longer than 40 yrs. 

%discrete

In the case of the discrete distance function (Eq.(\ref{discrete}) the mean value of distances between countries increases linearly with the time window size (Fig.\ref{fig:l1}). This observation is also supported by the value of the linear correlation coefficient (Eq.(\ref{korel})) which is very close to one (table \ref{corr}) for all of investigated structures. This relationship is caused by the properties of the applied distance function, (Eq.(\ref{discrete}), which accumulates the differences between the considered time series. Therefore it is not suggested to use it in order to compare properties of different time window sizes, unless properly normalised but it may be usefull in an analysis of the evolution of a system within a given time window size.

\begin{table}[h]
\begin{center}
\begin{tabular}{|c|c|}
\hline
& corr \\ \hline
LMST & 0.99854 \\ \hline
UMLP & 0.99717 \\ \hline
BMLP & 0.99712 \\ \hline
\end{tabular}
\caption{ \label{corr} Linear correlation coefficients between the time window and the mean distance beteew countries in the case of LMST, UMLP and BMLP.} 
\end{center}
\end{table}
As in the case of the statistical distance there are no significant differences between standard deviation, skewness, kurtosis for LMST, UMLP and BMLP (Fig.(\ref{fig:l1})).  The standard deviation has a maximum at 30 yrs time window. It means that  within this time window there is the largest spread of distances between considered countries. These results may be the most interesting ones for analysis, because the time evolution may reveal significant changes or a nontrivial evolution of the distances between countries. From an information content the point of view the LMST method gives the highest amount of information, because the standard deviation of mean distances is the highest for this graph algorithm.

%gauss

\begin{figure}
	\includegraphics[angle=-90,scale=0.25]{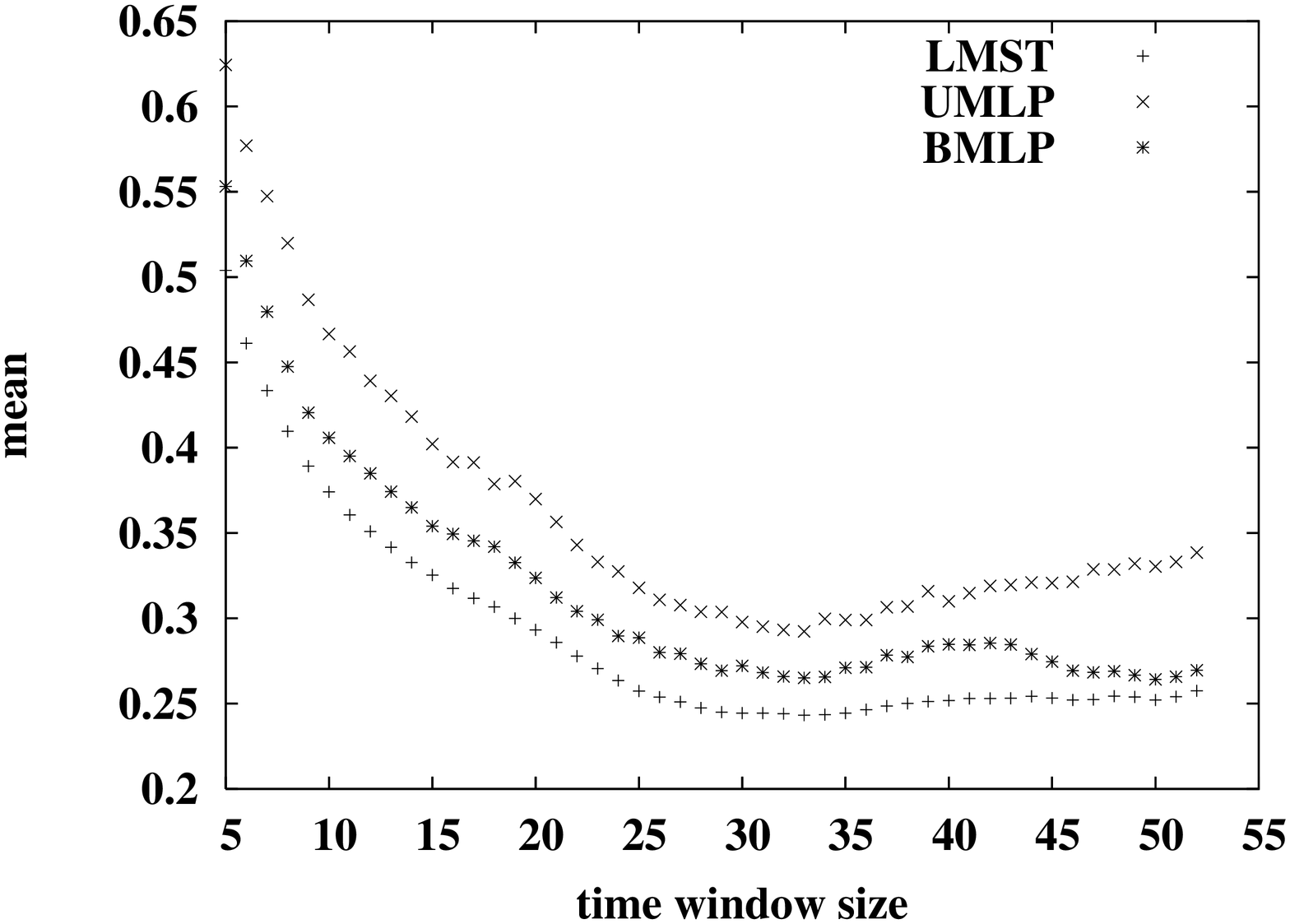}
	\includegraphics[angle=-90,scale=0.25]{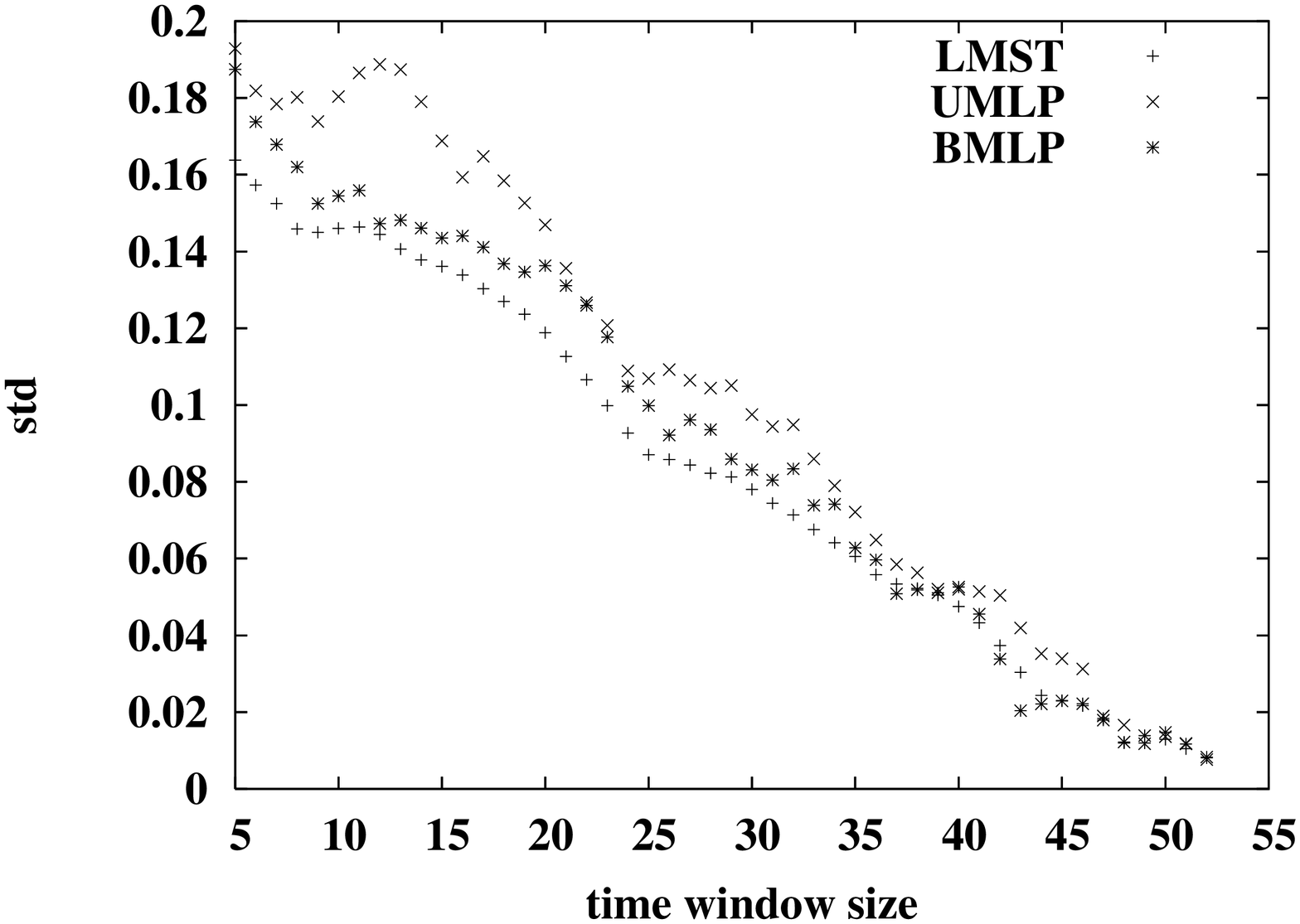} 
	\includegraphics[angle=-90,scale=0.25]{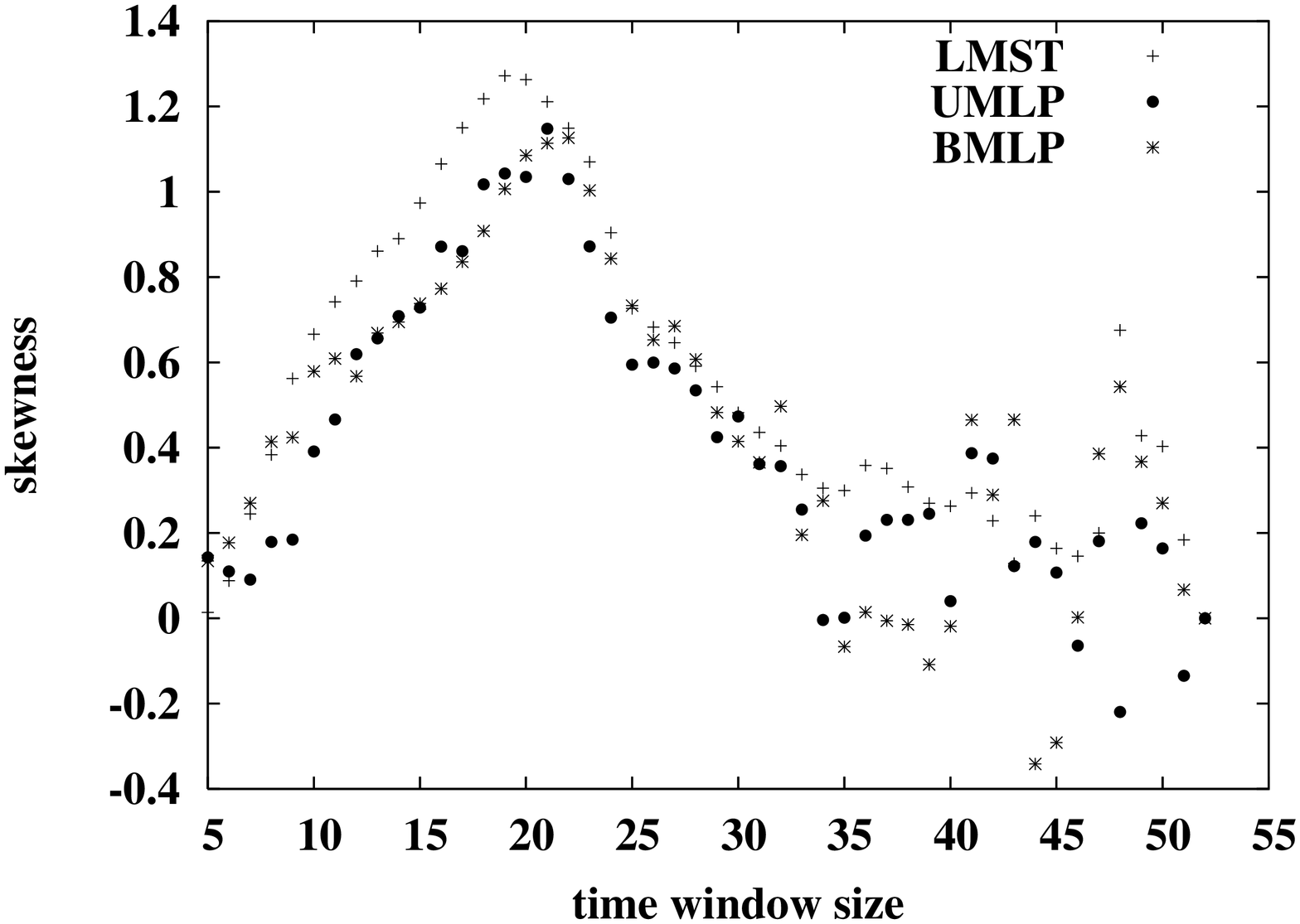}
	\includegraphics[angle=-90,scale=0.25]{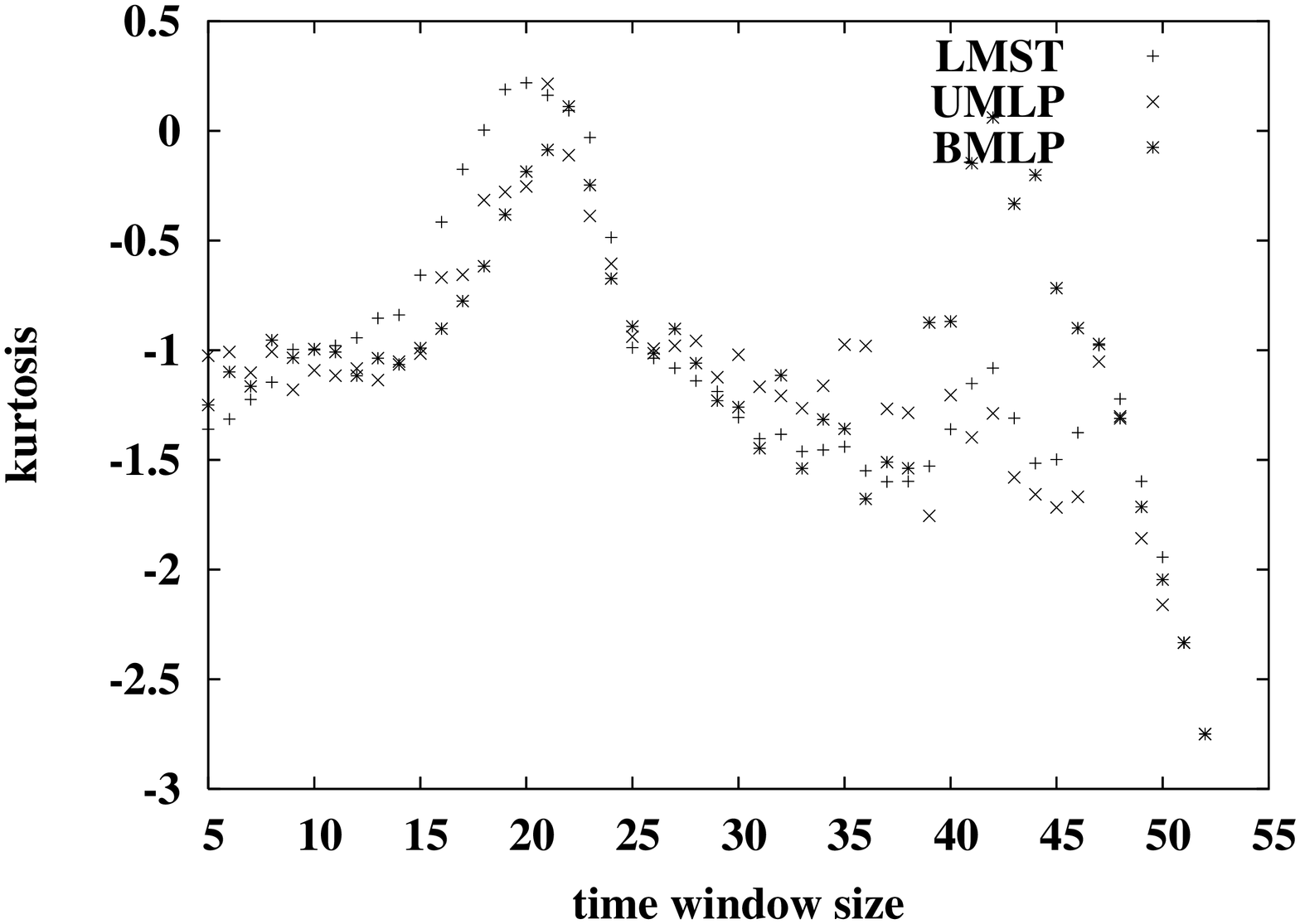}
	\caption{Statistical analysis of the graph properties obtained by application
of the distribution distance. The plots present mean value, standard deviation,
skewness and kurtosis of the distance between countries in the case of LMST, UMLP
and BMLP averaged over all G7 countries and the considered time interval as a
function of the time window size.} 	\label{fig:g1}
\end{figure}

In the case of the distribution distance defined by Eq.(\ref{distr_dist}), the mean distances between countries are decreasing monotonically up to the 30 yrs time window and stabilising or slowly increasing (with respect to the behaviour for the time window size < 30 yrs) for longer window size. The other statistical properties (standard deviation, skewness and kurtosis) are as in the previously considered cases (statistical distance and discrete distance) almost identical for all considered graph methods. The standard deviation is decreasing monotonically within the considered perion while skewness and kurtosis have a maximum for the time window size equal 20 yrs.

For all considered graph methods the LMST algorithm gives the lowest mean distance value (Figs \ref{fig:m}, \ref{fig:l1}, \ref{fig:g1}). It is also worth noticing that despite different distance metrics the graph methods give always the same order of functions. The lowest value is taken by the mean distance in the case of LMST, the second is BMLP and the highest value is received by application of the UMLP algorithm. This order is caused by an "optimisation level". In the case of the LMST a new point on the graph can be added at every graph node, in BMLP on one of both ends and in UMLP only on a one end so that the variety of possibilities is decreasing thereby resulting in less densly connected graph. Howether all function are very similar to each other. Since chain algorithms are significantly simpler numerically and the received information is similar therefore it may be usefull to apply one of the chain algorithms instead of building the MST or LMST tree.

\section{Conclusions}

The distance analysis consists in two steps. One is the choice of the distance metrics, the second is the graph analysis. It has been shown that the results do not depend significantly on the choice of the graph analysis, but rather on the distance function. The mean distance between G7 countries is increasing with the time window size in the case of the statistical and discrete distance, while the application of the distribution distance results in the opposite behaviour -- the mean distance between countries is decreasing with the time window size. In this situation it is extremely important to choose properly the distance function, because there are different advantages to each of the used distance functions. 
The first method Eq.(\ref{stat}), a statistical distance, is specially sensitive to linear correlations. 
The discrete Hilbert space $ L_q $ distance Eq.(\ref{discrete}) can be applied to any data and does not require any special properties of the data so this method seems to be very useful for comparing various sets of data. However the window size should not change in the analysis since it may influence significantly the results. 
The third method Eq.(\ref{distr_dist}) is the most sophisticated one since it requires a knowledge of the data distribution function, but then points out to similarities between data statistical properties. The main disadvantage of the last method is that it is sensitive to the size of the data set, since it requires fitting a distribution function. It is worth noticing that the results do not depend significantly on the choice of the graph analysis. Of course results may differ in details, but at the first stage it is useful to apply one of the chain methods (BMLP or UMLP) since they are much simpler than the LMST especially from the numerical point of view. Thus may help in the distance metrics choice. The optimal window size may be chosen on the analysis of the statistical properties of the appropriate structure \cite{bidirectional}. 

Various distance metrics have been investigated and new methods based on different distance metrics choice have been proposed in order to investigate the correlations between G7 countries. These methods of mean distance analyses could be also applied to stock market analysis.

\section{Acknowledgement}
This work is partially financially supported by FNRS convention FRFC 2.4590.01.
J. M. would like also to thank SUPRATECS for the welcome and hospitality.

\bibliographystyle{unsrt}
\bibliography{warsz.bib}

\begin{thebibliography}{1}

\bibitem{hurst}
M.~Couillard and M.~Davison.
\newblock A comment on measuring the {Hurst} exponent of financial time series.
\newblock {\em Physica A}, 348:404--418, 2005.

\bibitem{hursta}
A.~Carbone, G.~Castelli, and H.E. Stanley.
\newblock Time-dependent {Hurst} exponent in financial time series.
\newblock {\em Physica A}, 344:267--271, 2004.

\bibitem{window}
E.A. Maharaj.
\newblock Comparison of non-stationary time series in the frequency domain.
\newblock {\em Computational Statistics and Data Analysis}, 40:131--141, 2002.

\bibitem{dfa}
Z.~Chen, P.~Ch. Ivanov, K.~Hu, and H.~E. Stanley.
\newblock Effect of nonstationarities on detrended fluctuation analysis.
\newblock {\em Phys. Rev. E}, 65:041107, 2002.

\bibitem{probab}
M.A. Goldberg.
\newblock {\em An introduction to probability theory with statistical
  applications}.
\newblock Plenum Press, New York, 1984.

\bibitem{bidirectional}
M.~Ausloos and J.~Miskiewicz.
\newblock An attempt to observe economy globalization: the cross correlation
  distance clustering of the top 19 {GDP} countries.
\newblock {\em submitted for publication}, 2005.

\bibitem{analiza}
Maurin K.
\newblock {\em Analiza}.
\newblock PWN, Warszawa, 1991.

\end{thebibliography}

\end{document}